\documentclass[5p, times]{elsarticle}

\usepackage[colorlinks]{hyperref}
\usepackage[numbers]{natbib}
\usepackage{float}
\usepackage[nobiblatex]{xurl}
\usepackage{listings}
\usepackage{todonotes}
\usepackage{xcolor}
\colorlet{codebase}{yellow!5!}
\newcommand*{\code}{\lstinline[basicstyle=\ttfamily,breaklines=true, breakatwhitespace=false,prebreak=-]}
\makeatletter
\def\lst@lettertrue{\let\lst@ifletter\iffalse}
\makeatother

\usepackage{tikz,pgfplots}
\usetikzlibrary{automata,arrows}
\usetikzlibrary{calc,trees,positioning,arrows,chains,shapes}
\usetikzlibrary{shapes.geometric}
\usetikzlibrary{shapes.multipart}
\newtheorem{theorem}{Theorem}

\AtBeginDocument{%
  \hypersetup{%
    linkcolor=blue,%
    citecolor=red,%
  }%
}%

\usepackage{amsmath}

\newcommand{\pderiv}[2]{\ensuremath{\frac{\partial {#1}}{\partial {#2}}}}
\usepackage{amssymb}

\usepackage{todonotes}
\usepackage{comment}
\usepackage{soul}
\newcommand{\hp}[1]{\textcolor{magenta}{#1}}

\begin{document}

\newcommand{\hps}[1]{\hp{\st{#1}}}
\newcommand{\shk}[1]{\textcolor{blue}{#1}}
\newcommand{\ph}[1]{\textcolor{red}{#1}}

\journal{Joint Laboratory on Extreme Scale Computing Future Generation Computer Systems (JLESC-FGCS)}

\begin{frontmatter}

\title{MITgcm-AD v2: Open source tangent linear and adjoint modeling framework for the oceans and atmosphere enabled by the Automatic Differentiation tool Tapenade}

\author[inst1]{Shreyas Sunil Gaikwad}
\author[inst2]{Sri Hari Krishna Narayanan}
\author[inst3]{Laurent Hascoet}
\author[inst4]{Jean-Michel Campin}
\author[inst1]{Helen Pillar}
\author[inst1]{An Nguyen}
\author[inst2]{Jan Hückelheim}
\author[inst2]{Paul Hovland}
\author[inst1,inst5,inst6]{Patrick Heimbach}

\affiliation[inst1]{organization={Oden Institute for Computational Engineering and Sciences},
            addressline={The University of Texas at Austin}, 
            city={Austin},
            state={TX 78712},
            country={USA}}

\affiliation[inst2]{organization={Mathematics and Computer Science Division},
            addressline={Argonne National Laboratory}, 
            city={Lemont},
            state={IL 60439},
            country={USA}}

\affiliation[inst3]{organization={Institut National de Recherche en Informatique et Automatique},
            city={Valbonne 06902},
            country={France}}
            
\affiliation[inst4]{organization={Department of Earth, Atmospheric, and Planetary Sciences},
            addressline={Massachusetts Institute of Technology}, 
            city={Boston},
            state={MA 02139},
            country={USA}}
            
\affiliation[inst5]{organization={Jackson School of Geosciences},
            addressline={The University of Texas at Austin}, 
            city={Austin},
            state={TX 78712},
            country={USA}}

\affiliation[inst6]{organization={Institute for Geophysics},
            addressline={The University of Texas at Austin}, 
            city={Austin},
            state={TX 78712},
            country={USA}}

\begin{abstract}
The Massachusetts Institute of Technology General Circulation Model (MITgcm) is widely used by the climate science community to simulate planetary atmosphere and ocean circulations. A defining feature of the MITgcm is that it has been developed to be compatible with an algorithmic differentiation (AD) tool, TAF, enabling the generation of tangent-linear and adjoint models. These provide gradient information which enables dynamics-based sensitivity and attribution studies, state and parameter estimation, and rigorous uncertainty quantification. Importantly, gradient information is essential for computing comprehensive sensitivities and performing efficient large-scale data assimilation, ensuring that observations collected from satellites and in-situ measuring instruments can be effectively used to optimize a large uncertain control space. As a result, the MITgcm forms the dynamical core of a key data assimilation product employed by the physical oceanography research community: Estimating the Circulation and Climate of the Ocean (ECCO) state estimate. Although MITgcm and ECCO are used extensively within the research community, the AD tool TAF is proprietary and hence inaccessible to a large proportion of these users. 

The new version 2 (MITgcm-AD v2) framework introduced here is based on the source-to-source AD tool Tapenade, which has recently been open-sourced. Another feature of Tapenade is that it stores required variables by default (instead of recomputing them) which simplifies the implementation of efficient, AD-compatible code.
The framework has been integrated with the MITgcm model’s main branch and is now freely available.
\end{abstract}


\end{frontmatter}

\section{Introduction}\label{sec:Introduction}

\noindent
Mathematical modeling of geophysical fluids is a complex undertaking that necessarily involves several approximations (constitutive models, spatial discretization, and subgrid-scale parameterizations) to close a system of high-fidelity equations such as conservation of mass, momentum, energy, and tracers. Examples of such parameters include those used to represent aerosol and cloud microphysics in the atmosphere, the coefficients of the mixing parameterizations in the ocean, and the basal sliding coefficients below ice sheets. Model boundary and initial conditions are also required and often poorly constrained. Meaningful interpretation of model output therefore demands investigation of the impact of these uncertain parameters, initial conditions, and boundary conditions on the simulated state, and an effort to identify their ``best" values for some specific metric. The traditional (brute-force) investigation entails running large forward ensembles, applying separate perturbations to different uncertain inputs (or ``controls"), and then quantifying the spread in the simulated state (e.g., \citep{Kay.2015,Palmer.2019,Deser:2020bj,Lehner.2020}). 
Appropriate values for the uncertain controls may then be chosen as those providing the best model validations. For complex models with a high-dimensional control space, however, truly comprehensive sensitivity investigation and rigorous model calibration are not computationally tractable with this approach.

\subsection{The Adjoint Method}
Alternatively, the adjoint method can provide accurate sensitivity information comprehensively and efficiently. 
While tangent linear and adjoint operators are more general mathematical objects, in the context of a 
model, we assume that we work in a space of discrete, i.e., finite-dimensional spaces. The model is nonlinear and maps our discrete input parameters $\mathbf{u} \in \mathbb{R}^N$ (an aggregated vector that may include spatially distributed model parameters, initial and boundary conditions, etc.) to the ocean state $\mathbf{x} \in \mathbb{R}^M$. An ``objective function" $\mathcal{J}$ then projects the state $\mathbf{x}(\mathbf{u})$ (or some diagnostic thereof) to some scalar quantity of interest. 
This objective function can either be a model-data misfit that we seek to reduce through optimization or a scalar quantity such as some ocean transport (e.g., of volume, heat, or salt through a particular strait) whose variations or sensitivity we wish to understand with respect to our controls. The model $\mathcal{M}$ and the objective function $\mathcal{J}$ can thus be defined as the following mappings: 

\begin{equation}
        \mathcal{M} : \, \mathbb{R}^N \ni \mathbf{u} \, \rightarrow \,
        \mathcal{M}(\mathbf{u})\, = \, \mathbf{x} \in \mathbb{R}^M,
\end{equation}
and
\begin{equation}
        \mathcal{J} : \, \mathbb{R}^N \ni \mathbf{u} \, \rightarrow \,
        \mathcal{J}(\mathcal{M}(\mathbf{u}))\,=\,\mathcal{J}(\mathbf{x}) \in \mathbb{R}.
\end{equation}

Consider the time-dependent problem, where evaluating the model over the
time window $t\in(t_0,t_f)$ involves successively stepping the nonlinear system of
equations $\mathcal{L}$ forward in time as
\begin{equation}
        \mathbf{x}(t_f) = \mathcal{M}(\mathbf{u}) =
        \mathcal{L}_{N_t}\left(\cdots \mathcal{L}_0(\mathbf{u})\right) \,
        , \label{eq:time_depedent_problem}
\end{equation}
where 
$t_f \, = \, t_0 \, + \, N_t \, \Delta t$,
and $\mathcal{L}_n$ maps the state at time step $n-1$ to $n$, such that
$\mathbf{x}(t_{n}) \, = \, \mathcal{L}_n\left(\mathbf{x}(t_{n-1})\right)$ and 
$\mathbf{x}(t_0) \, = \, \mathcal{L}_0(\mathbf{u})$. In this case, it is thus implicitly assumed that only initial conditions are a part of the control space and not parameter and boundary conditions. A more general consideration of the control space is possible, e.g. \citep{wunsch2007}.

Now consider linear perturbations in the cost function arising from perturbations in the input parameters using the Taylor expansion, and writing the scalar product representing the directional derivative $(\partial \mathcal{J}/{\partial \mathbf{u}})^T \, \cdot \, \delta\mathbf{u}$ as bilinear form $< . \, , \, . >$, we have: 
\begin{equation*}
        \mathcal{J}(\mathbf{u}_0 + \delta\mathbf{u}) = \mathcal{J}(\mathbf{u}_0) + \delta \mathcal{J} + 
        \mathcal{O}\left(\delta \mathcal{J}^2 \right),
\end{equation*}
where 
\begin{equation}
\begin{aligned}
        \delta \mathcal{J} &= \left\langle \pderiv{\mathcal{J}}{\mathbf{u}}\,,\,\delta\mathbf{u}\right\rangle \\
                   &=\left \langle \pderiv{\mathcal{J}}{\mathbf{x}(t_1)}\,,\,
        \pderiv{\mathbf{x}(t_1)}{\mathbf{u}}\delta\mathbf{u}\right\rangle \\
        &= \left\langle \pderiv{\mathcal{J}}{\mathbf{x}(t_2)}\,,\,\mathbf{L}_1 
        \pderiv{\mathbf{x}(t_1)}{\mathbf{u}}\delta\mathbf{u}\right\rangle  \\
        &= \left\langle \pderiv{\mathcal{J}}{\mathbf{x}(t_3)}\,,\,\mathbf{L}_2 \mathbf{L}_1
        \pderiv{\mathbf{x}(t_1)}{\mathbf{u}}\delta\mathbf{u}\right\rangle  \\
        &= \left\langle \pderiv{\mathcal{J}}{\mathbf{x}(t_f)}\,,\,\mathbf{L}_{N_t-1} \cdots \mathbf{L}_1
        \pderiv{\mathbf{x}(t_1)}{\mathbf{u}}\delta\mathbf{u}\right\rangle  \, .
        \label{eq:general_tlm_eq}
\end{aligned}
\end{equation}
$\mathbf{L}_n = \pderiv{\mathbf{x}(t_n+1)}{\mathbf{x}(t_n)}$ is the \textit{tangent
linear model}, a linearization of $\mathcal{L}$ at time $t_n$ about $\mathbf{u}_0$.
Now consider the model adjoint $\mathbf{L}_n^T$
\begin{equation}
\begin{aligned}
        \delta \mathcal{J} &= 
        \left\langle \pderiv{\mathbf{x}(t_1)}{\mathbf{u}}^T \mathbf{L}_1^T \cdots
                \mathbf{L}_{N_t-1}^T \pderiv{\mathcal{J}}{\mathbf{x}(t_f)}\,,\,
        \delta\mathbf{u}\right\rangle  \\
        &= \left\langle \pderiv{\mathcal{J}}{\mathbf{u}}\,,\,\delta\mathbf{u}\right\rangle \, .
        \label{eq:general_adjoint_eq}
\end{aligned}
\end{equation}
A closer look at the dependencies in Equation (\ref{eq:general_adjoint_eq}) reveals the following,
\begin{equation}
\begin{aligned}
        \delta \mathcal{J} &= 
        \left\langle \pderiv{\mathbf{x}(t_1)}{\mathbf{u}}^T \mathbf{L}_1^T (\mathbf{x}(t_0))\cdots
                \mathbf{L}_{N_t-1}^T(\mathbf{x}(t_{N_t-1})) \pderiv{\mathcal{J}}{\mathbf{x}(t_f)}\,,\,
        \delta\mathbf{u}\right\rangle \,
        \label{eq:general_adjoint_eq_2}
\end{aligned}
\end{equation}
Whereas the scalar product involving tangent linear operators $\mathbf{L}_n$ in (\ref{eq:general_tlm_eq}) amounts to a directional derivative, mapping from $\mathbb{R}^N$ to $\mathbb{R}^1$, the scalar product involving the adjoint operators $\mathbf{L}_n^T$ in (\ref{eq:general_adjoint_eq_2}) provides the gradient and maps from $\mathbb{R}^1$ to $\mathbb{R}^N$.
 Adjoints can also be used to approximate higher-order derivative information during optimization
 for accelerating convergence and quantifying uncertainty \citep{Chen2011, Petra2012, Alghamdi2020}.  

We are interested in using the adjoint method with MITgcm, a finite volume numerical model that is used to study the evolution of atmospheric and oceanic flows by solving the Boussinesq approximation of the Navier-Stokes equations on the rotating sphere in either hydrostatic or non-hydrostatic form \citep{Marshall1997,Marshall1998}. Here we consider hydrostatic ocean-only configurations of the model. The MITgcm comprises $\mathcal{O}\left(10^6\right)$ lines of code and typical simulations may span millions of iterative solves for velocity, temperature, and salinity. Each of these timesteps, in turn, includes the execution of an iterative solver inverting for pressure. Horizontal domain decomposition, accompanied by an application-specific communication library \citep[][]{Hoe99}, 
enables efficient and scalable parallel processing.

\subsection{Automatic Differentiation}
Generating and maintaining adjoint models manually for code bases such as MITgcm is a cumbersome and error-prone activity. One can instead use 
automatic/algorithmic differentiation (AD), also known as autodiff. AD is a method to compute the derivatives of source code~\cite{Griewank2008EDP,NaumannBook}. 
It relies on viewing any numerical model as a composition of operations evaluated in a certain order. AD exploits the chain rule of calculus to compute the derivative of this entire composition one operation at a time. AD is advantageous in sparing user commitment of updating the adjoint manually with every forward model development \cite{Heimbach2005}.

When differentiating a large application, however, the application of an AD tool is beset by challenges that are not often encountered in simpler straight-line codes. These include:
\begin{itemize}
 \item Adapting the application’s build system to include differentiation. Often the code that is needed to run a case of interest forms only a subset of the entire code base. Tools such as makefiles and scripts as well as preprocessor directives can be used to select the exact lines of code that are needed for differentiation. So the AD tool needs to be integrated closely with the model.  
 \item The parallelism methodology of the code should be differentiable by the AD tool (unless custom rules are provided that enable bypassing the corresponding code). 
 Codes may use MPI~\cite{gropp1999using}, OpenMP~\cite{chandra2001parallel}, RAJA~\cite{beckingsale2019performance}, GPUs, etc. Different AD tools have differing levels of support for these constructs and the programming languages that use them. 
Tapenade supports direct differentiation of OpenMP parallel work-sharing loops in
C/C\texttt{++} and Fortran~\cite{10.1145/3472796} and a subset of MPI primitives~\cite{ampi}. TAF offers support for OpenMP and MPI.
The Enzyme automatic differentiation framework~\cite{enzymeNeurips}, built using the LLVM compiler infrastructure, supports synthesizing adjoints of GPU kernels ~\cite{enzymeGPU} as well as MPI, OpenMP, and Julia Tasks~\cite{10.5555/3571885.3571964}. 
The Fortran frontend for LLVM, however, is still undergoing development.

 \item The sequence of operations to which the chain rule is applied may be interrupted by I/O operations causing the AD tool to conclude the output of the model does not depend on the input.
 \item The model may rely on library routines whose source code may not be available or is not amenable to AD. This may require users to supply derivative code for the routines to the AD tool.
 \item The model may call iterative solvers that, if differentiated directly, will produce inaccurate and inefficient derivatives.
 \item During an adjoint run of a model containing time-stepping loops, the states are needed in reverse of the sequence they are computed in the forward run. This can cause the adjoint model to run out of memory and must be alleviated by checkpointing.
 \item It is not easy to know whether the derivatives that have been obtained are correct and how to debug sources of error. Commonly used methods to check correctness include comparison of derivatives computed using another tool, and comparison against derivatives computed by the finite difference approximation. Other approaches include the dot product test to check for internal consistency of the forward and reverse mode and the Taylor remainder convergence test. 

\end{itemize}

\subsection{AD in MITgcm}
A defining aspect of the MITgcm is that it has been developed to be compatible with an AD tool, namely the Tangent Linear and Adjoint Model Compiler (TAMC) and its successor TAF \citep{Giering1998}. Obtaining computationally tractable adjoint calculations required significant code modifications.
As a result of this effort, the MITgcm has emerged as a powerful tool for extensive investigation of ocean sensitivity \citep[e.g.,][]{Marotzke1999,Heimbach2011,Pillar2016,Smith2019,Nguyen2020} and adjoint-based parameter and state estimation \citep[e.g.,][]{wunsch2007,Wunsch2013,Forget2015,Nguyen2021}.
AD has been essential in the development of NASA's Estimating the Circulation and Climate of the Ocean (ECCO) state and parameter estimation framework, enabling O(10$^9$) ocean and sea-ice observations from satellite and in-situ platforms to be assimilated into a multi-decadal integration of the MITgcm to constrain the simulated state through adjoint-based adjustment of O(10$^7$) controls 
\cite{Stammer.2002,wunsch2007,Heimbach:2019dn}.

However, the MITgcm adjoint is not widely available because the currently used AD tool, TAF is proprietary. 
Development of an open-source AD tool for complex general circulation models is essential to make adjoints and their powerful applications for climate science available to the wider research community, as well as easy to use through increased transparency \citep{Utke2008}. Previously, the open source AD tool OpenAD has been used for a limited set of experiments within the MITgcm \citep{Naumann2006, Utke2008, gmd-9-1891-2016, https://doi.org/10.1029/2020JC016370}.

\subsection{Contributions}
Here we describe work undertaken to differentiate MITgcm using Tapenade, an open-source Automatic Differentiation Engine \cite{Hascoet2013}. 
Tapenade can be used to differentiate Fortran77, Fortran90, Fortran95, or C codes, with partial extensions to parallel dialects such as MPI, OpenMP, or Cuda.
Previous applications of Tapenade include differentiation of a global sea-ice model \citep{Kim2006SAa}, CFD Solver Development \citep{Praveen2006Acd, Giles2006Uad}, parameter estimation for a river hydraulics model and adjoint sensitivity analysis of rainfall-runoff \citep{Castaings2005ADA}, and optimal design analysis for a supersonic plane \citep{Courty2003Rad}. MITgcm-AD v2 represents significant progress in the use of Tapenade.

 The remainder of the paper is organized as follows. Our work to develop compatibility between the MITgcm and Tapenade is described in section \ref{section:impement_ad}, which includes a discussion of the challenges faced and modifications made in the main trunk of the MITgcm. We demonstrate the application of Tapenade to global ocean configurations of the MITgcm in section \ref{section:example}, comparing performance to that obtained by employing TAF. Our conclusions and directions for future research are given in section \ref{section:conclusion}. Instructions for reproducing our results using MITgcm-AD v2 are given in \ref{section:build_ad}.

\section{AD development }
\label{section:impement_ad}
\noindent
Our approach is based on prior experience in replacing the use of OpenAD \citep{gmd-13-1845-2020} to differentiate the ice sheet model SICOPOLIS \citep{Greve1997a, Greve1997b} with Tapenade \citep{Gaikwad2023}. Both OpenAD and TAF were already applied in MITgcm through modifications to MITgcm's compilation process. We undertook the following modifications to apply Tapenade similarly, i.e., to integrate Tapenade into the build process for differentiating and then compiling the original and differentiated code: 
\begin{itemize}
\item We edited the \code{genmake2} utility script to generate a Makefile that invokes Tapenade and recognizes relevant file dependencies. \code{genmake2} plays a role similar to \code{autotools} for the MITgcm, making it easy to run the model on different platforms and hardware architectures. 
Tapenade may find that some files are not involved in the differentiable chain of the computation, and therefore no corresponding differentiated file is created. Unfortunately, this so-called \textit{activity analysis} depends on the considered MITgcm test case, and the set of differentiated files cannot be predicted. In order to keep the Makefile generic, we create empty versions of all possible differentiated files prior to differentiation, and let the AD tool overwrite the ones that must hold differentiated code.
\item We provided subroutines to handle file I/O within the computation. Users are typically interested in differentiating a particular subset of the MITgcm code and computing the derivatives of identified output variables w.r.t identified input variables. These variables are called \textit{dependent} and \textit{independent} variables, respectively. Variables that are affected by independent variables and in turn affect the dependent variables are called {\em active variables}. For efficient derivative computation, Tapenade employs activity analysis to identify active variables and computes derivatives only for these variables. Several operations in the primal computation, however, can cause this analysis to fail. One of these is file I/O where a variable that is intended to be active is written to a file and later read from the same file. In MITgcm this occurs mainly when
(i) accumulating state variables for evaluating model-data misfit cost function terms after completion of the model integration (e.g., daily-mean sea surface height fields, vertical water column-wise daily hydrographic profiles, monthly mean ocean bottom pressure anomalies based on time-step snapshots);
(ii) accumulating time-varying control variable fields with flexible averaging periods (e.g., accumulation of control variables for time-varying adjustments of either air-sea fluxes or the atmospheric surface state itself, with adjustment intervals defined as multiples of the model time step -- such as daily, weekly, biweekly, etc.).
Because file I/O is seen as a passive operation, without any intervention activity analysis will determine that the variable is itself passive.  

To overcome this problem in the MITgcm, we identified code that performs file I/O of variables that should be active. For example, within the routine 
\code{WRITE_REC_XY_RL} that acts as an interface to the low-level I/O, consider a subroutine \code{ACTIVE_WRITE_XY}. We provided directives to Tapenade identifying the properties of the activity and interdependencies of the arguments of this subroutine:

\begin{lstlisting}[language = fortran]
subroutine active_write_xy:
  external:
  shape: (par 1, par 2, par 3, par 4, par 5, par 6)
  type: (none(),none(),none(),none(),none(),none())
  NotReadNotWritten:  (0,0,0,0,0,0)
  ReadNotWritten:     (1,1,1,1,1,1)
  NotReadThenWritten: (0,0,0,0,0,1)
  ReadThenWritten:    (0,0,0,0,0,1)
\end{lstlisting}

When provided with flow information for a subroutine, Tapenade will 
not differentiate through the subroutine, but will simply use the information to carry out activity analysis correctly. The file \code{tools/TAP_SUPPORT/flow_tap} contains all the flow directives for such routines.

The user must provide, for \code{ACTIVE_WRITE_XY}, appropriate subroutines, in this case, \code{ACTIVE_WRITE_XY_D} and \code{ACTIVE_WRITE_XY_B} that compute the derivatives in tangent linear and adjoint modes, respectively. In the case of file I/O where no computation takes place, the custom routines simply write the intermediate derivative values to the file or read the intermediate derivative values from a file to initialize or modify the derivatives of some active variable.

Similar steps are needed for the equivalent read subroutines, in this case \code{ACTIVE_READ_XY}.

The subroutines for all such file I/O subroutines are in 
the files \code{pkg/tapenade/active_read_tap.F} and \\
\code{pkg/tapenade/active_write_tap.F}.

\item  
To minimize communication between tiles in domain-decomposed parallel execution of the model, each tile consists of an interior portion and a ``halo region'' which carries information about the neighboring tile. The size of the halo region depends on the computational stencil of the numerical schemes selected. 
When needed, exchange routines will copy the information of each tile's interior to the halo region of its neighbors. These communications are managed via MPI.
The purpose of the halo region is to reduce such communications to a minimum within each time step.
While we could have attempted to utilize Tapenade's experimental adjoinable MPI capability for differentiating through the MPI communication, we instead provide Tapenade with handwritten, reverse-mode exchange subroutines for the adjoint. Fortunately, this only involves writing wrappers around the subroutines already developed for AD with TAF \citep{Heimbach2005}.
 Importantly, the exchange routines written for MITgcm are agnostic of the specific MPI library used (e.g., OpenMPI, MPICH, Intel MPI, etc.). This will be handled by low-level interface routines), and so the derivative code too will be agnostic to the library.

\item We wrote subroutines to handle the differentiation of iterative solvers.
The pressure method used in the MITgcm consists of solving a 2-D (or 3-D) elliptic problem which guarantees that the continuity equation is satisfied. This is achieved via an iterative, preconditioned conjugate gradient solver
~\cite{Marshall1997} for computing a system of linear equations:
\begin{equation}
\begin{aligned}
    A \cdot x = b \rightarrow x := \mathrm{solve}(A,b)
        \label{eq:iterative_solve_eq}
\end{aligned}
\end{equation}
AD tools should not differentiate through iterative solvers because it reduces the efficiency and accuracy of the code. Furthermore, the solver code may be provided by external packages and not exposed to the AD tool. The adjoints $\bar{A}$ and $\bar{b}$ can instead be computed from $\bar{x}$ using the original solver call itself~\cite{Giles2008}.

\begin{equation}
\begin{aligned}
A^T \cdot \bar{b} = \bar{x} \rightarrow \bar{b} := \mathrm{solve}(A^T,\bar{x})  \\
\bar{A} := -x^T \cdot \bar{b}  \label{eq:iterative_solve_eq_ad}
\end{aligned}
\end{equation}

For a solver call in MITgcm such as \code{CALL solver(A,y,x)}, we have defined the pseudo-code of the adjoint solver call\\ \code{CALL solver_B(A, Ab, y, by, x, xb)} by calling \code{solver} twice to solve Eqs (\ref{eq:iterative_solve_eq_ad}), without using Tapenade to differentiate through any external library or fixed point iterations, essentially treating the solver as a black box.

\item We added checkpointing support to enable long time integration of the adjoint model. From Eq.~(\ref{eq:general_adjoint_eq_2}), the computation of the adjoint of step $\mathbf{L}_i^T$ is dependent on the computed state $\mathbf{x}(t_i)$ and the adjoint state $\mathbf{L}_{i+1}^T$. The sequence of timesteps in the adjoint computation is $\mathbf{L}_N^T$ down to $\mathbf{L}_0^T$. This implies, in a naive setting, that for computing $\mathbf{L}_i^T$, the states $\mathbf{x}(t_0),\,\mathbf{x}(t_1),\,\dots , \, \mathbf{x}(t_{i})$ will be stored in memory. 
Typical applications of the MITgcm to interrogate seasonal to multidecadal ocean variability will consist of O(10$^4$) timesteps, each requiring O(10$^8$) bytes, which is prohibitively large to be held in memory. Because only the forward state $\mathbf{x}(t_{i})$ is needed to compute $\mathbf{L}_i^T$, however, AD tools support storing only a subset of the states $\mathbf{x}(t_0)\rightarrow \mathbf{x}(t_{i})$ with the rest being recomputed from the stored values. This approach is commonly known as checkpointing and offers a tradeoff between the recomputation of states and their storage. 

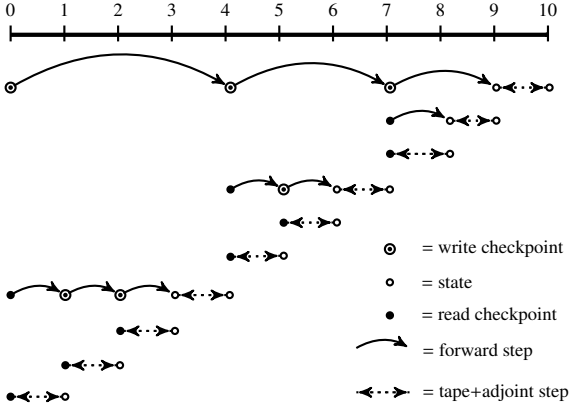
\begin{figure}[t!]
  \begin{center}
  \begin{tikzpicture}[>=stealth',shorten >=1pt,thick,black,auto, scale = 0.7, transform shape,initial/.style={black,fill=black,text=white}]
\tikzset{every state/.style={minimum size=0pt, minimum width=0pt, minimum height=0pt,scale=.35,black}}
  \foreach \x in {0,1,2,...,10}
    {        
      \coordinate (A\x) at ($(0,0)+(\x*1.01cm,0)$) {};
      \draw ($(A\x)+(0,5pt)$) -- ($(A\x)-(0,5pt)$);
      \node at ($(A\x)+(0,3ex)$) {\x};
    }
  \draw[ultra thick] (A0) -- (A10) -- ($(A10)+0.01*(1,0)$);
  
\node[state,accepting]         (A)    at (0,-1)                 {};
\node[state,accepting]         (B) [right=4cm of A]             {};
\node[state,accepting]         (C) [right=7cm of A]             {};
\node[state]                   (D) [right=9cm of A]            {};
\node[state]                   (E) [right=10cm of A]            {};
\node[state,fill=black]         (C1) [below=0.5cm of C]           {};
\node[state]                   (D1) [below=0.5cm of D]           {};
\node[state]                   (F) [right=1cm of C1]            {};
\node[state,fill=black]         (C2) [below=0.5cm of C1]           {};
\node[state]                   (F1) [below=0.5cm of F]            {};
\node[state,fill=black]         (B1) [below=1.8cm of B]             {};
\node[state,accepting]         (G) [right=0.87cm of B1]             {};
\node[state]                   (H)  [right=0.87cm of G]            {};
\node[state]         (C3) [right=0.87cm of H]           {};
\node[state,fill=black]         (G1) [below=0.5cm of G]             {};
\node[state]                   (H1)  [below=0.5cm of H]            {};
\node[state,fill=black]         (B2) [below=1.15cm of B1]             {};
\node[state]         (G2) [below=0.5cm of G1]             {};
\node[state,fill=black]         (A1) [below=3.8cm of A]                     {};
\node[state,accepting]         (I) [right=0.9cm of A1]                     {};
\node[state,accepting]         (J) [right=0.9cm of I]                     {};
\node[state]                   (K) [right=0.9cm of J]                     {};
\node[state]         (B3) [right=0.9cm of K]             {};
\node[state,fill=black]         (J1) [below=0.55cm of J]                     {};
\node[state]                   (K1) [right=0.9cm of J1]                     {};
\node[state,fill=black]         (I1) [below=1.2cm of I]                     {};
\node[state]         (J2) [right=0.9cm of I1]                     {};
\node[state,fill=black]         (A2) [below=1.8cm of A1]                     {};
\node[state]         (I2) [right=0.9cm of A2]                     {};
\node[state,accepting]         (C4) [below=1cm of C3,label={}]    {};
\node[]                        (C5) [right=0.4cm of C4]    {= write checkpoint};
\node[state]                   (C6) [below=0.5cm of C4,label={}]    {};
\node[]                          (C16) [right=0.4cm of C6]    {= state};
\node[state,fill=black]   (C7) [below=0.5cm of C6,label={}]    {};
\node[]                        (C8) [right=0.4cm of C7]    {= read checkpoint};
\node[]                        (C9) [below=0.5cm of C7,label={}]    {};
\node[]                        (C10) [left=0.5cm of C9,label={}]    {};
\node[]                        (C11) [right=1cm of C10,label={}]    {};
\node[]                        (C12) [right=-0.15cm of C11]    {= forward step};
\node[]                        (C13) [below=0.5cm of C10,label={}]    {};
\node[]                        (C14) [below=0.5cm of C11,label={}]    {};
\node[]                        (C15) [right=-0.15cm of C14]    {= tape+adjoint step};

\path[->] (C10) [bend left] edge node [right] {}  (C11);
\path[->] (C13) edge[dotted] node [right] {}  (C14);
\path[->] (C14) edge[dotted] node [right] {}  (C13);

\path[->] (A) edge [bend left] node [right] {}  (B)
       (B) edge [bend left] node [right] {}  (C)
       (C) edge [bend left] node [right] {}  (D);
\path[->] (D) edge[dotted] node [right] {}  (E);
\path[->] (E) edge[dotted] node [right] {}  (D);
\path[->] (C1) edge [bend left] node [right] {}  (F);
\path[->] (F) edge[dotted] node [right] {}  (D1);
\path[->] (D1) edge[dotted] node [right] {}  (F);
\path[->] (C2) edge[dotted] node [right] {}  (F1);
\path[->] (F1) edge[dotted] node [right] {}  (C2);
\path[->] (B1) edge [bend left] node [right] {}  (G);
\path[->] (G) edge [bend left] node [right] {}  (H);
\path[->] (H) edge[dotted] node [right] {}  (C3);
\path[->] (C3) edge[dotted] node [right] {}  (H);
\path[->] (G1) edge[dotted] node [right] {}  (H1);
\path[->] (H1) edge[dotted] node [right] {}  (G1);
\path[->] (B2) edge[dotted] node [right] {}  (G2);
\path[->] (G2) edge[dotted] node [right] {}  (B2);
\path[->] (A1) edge [bend left] node [right] {}  (I);
\path[->] (I) edge [bend left] node [right] {}  (J);
\path[->] (J) edge [bend left] node [right] {}  (K);
\path[->] (K) edge[dotted] node [right] {}  (B3);
\path[->] (B3) edge[dotted] node [right] {}  (K);
\path[->] (J1) edge[dotted] node [right] {}  (K1);
\path[->] (K1) edge[dotted] node [right] {}  (J1);
\path[->] (I1) edge[dotted] node [right] {}  (J2);
\path[->] (J2) edge[dotted] node [right] {}  (I1);
\path[->] (A2) edge[dotted] node [right] {}  (I2);
\path[->] (I2) edge[dotted] node [right] {}  (A2);
\end{tikzpicture}
  \end{center}
\caption{Top: Binomial checkpointing schedule for $l=10$ time steps and $c=3$ checkpoints. First, state $0$ is stored which serves as the input to time step $1$. After 4 time steps, state $4$ (input to time step $5$) is stored, followed by state $7$ (input to time step $8$). After time steps $8$ and $9$ are executed, the adjoint step $10$ is computed. Following this, the previously stored state $7$ is restored and time step $8$ is executed once more followed by adjoint step $9$. The rest of the schedule follows similarly~\cite{doi:10.1137/1.9781611976229.3}.}
	\label{fig:binomialcheckpointing}
\end{figure}

Several checkpointing techniques exist, including periodic, multilevel, and binomial checkpointing. 
For multi-annual integrations of MITgcm, the approach followed with TAF is a static $N$-level checkpointing (with the default $N=3$, see \cite{Heimbach2005}). Tapenade instead implements the approach known as binomial checkpointing \cite{Griewank1992ALG,revolve} which is based on the following theorem.  
\begin{theorem}
  \label{theo:bino}
Let $c$ be the number of available checkpoints and $l=N_t$ the number  of
time steps. The minimal number of time steps evaluated during the adjoint calculation 
given by 
	\begin{align*}
	t(l,c) &= \min_{1 \leq \hat{l} < l} \{\hat{l}\!+t(\hat{l},c)\!+\!t(l\!-\!\hat{l},c\!-\!1) \} \; \forall l>1,c>1,\\
	t(1,c) &= 0\;\;  \forall c > 0,\qquad 	 	t(l,1) = \frac{(l-1)l}{2}\;\; \forall l > 0
	\end{align*}
	has the explicit solution
	\begin{align} 
	 	t(l,c) = rl - \beta(c+1,r-1), \label{binom_formular}
	\end{align}
	where $\beta(c,r)= \binom{c+r}{c}$ and the repetition number
        $r$  is the unique integer, such that
	 \begin{align*} 
	 	\beta(c,r-1) < l \leq \beta(c,r). 
	 \end{align*}
\end{theorem}

This theorem allows a binomial checkpointing schedule to be created that is optimal in the number of recomputations performed. Figure~\ref{fig:binomialcheckpointing} shows an example of binomial checkpointing where $l=10$ and $c=3$. Binomial checkpointing is built-in in Tapenade, requiring no modification of the time loop other than adding one preprocessing directive.

\end{itemize}

\begin{table}[H] 
\caption{Verification experiments and tutorials (as named in the MITgcm) which exercise Tapenade-generated adjoint and tangent-linear models for a variety of dependent and independent variables.}
\label{table:compatible_verification_exps}
\centering{
\begin{tabular}{|c|c|}
\hline
tutorial\_tracer\_adjsens & OpenAD \\ 
\hline
tutorial\_global\_oce\_biogeo & global\_with\_exf\\
\hline
isomip&halfpipe\_streamice\\
\hline
global\_ocean.cs32x15&lab\_sea\\
\hline
tutorial\_barotropic\_gyre&tutorial\_baroclinic\_gyre\\
\hline
global\_oce\_biogeo\_bling&\\
\hline
\end{tabular}}\\
\end{table}

The changes above allowed us to differentiate most official tutorial and verification experiments (also known as regression or continuous integration tests) that already use AD with TAF (for MITgcm, ``tutorials'' refer to regression tests that are documented in some detail in MITgcm's online manual (see \url{https://mitgcm.readthedocs.io})). Tapenade is currently compatible with 10 tutorial/verification experiments (Table \ref{table:compatible_verification_exps}) and 25 packages (Table \ref{table:compatible_pkg}), including packages that handle sea ice and biogeochemistry in the ocean. As such, it far exceeds the number of packages that OpenAD was able to handle.  It is also compatible with NetCDF I/O (in the sense that Tapenade will properly handle calls to NetCDF library routines). We have added a list of checks to the MITgcm unit tests to track whether changes to the MITgcm code break the AD process using Tapenade.

\begin{table}[H] 
\caption{MITgcm packages for which Tapenade produces proper derivative code.}
\label{table:compatible_pkg}
\small
\centering{
\begin{tabular}{|c|c|c|c|c|c|}
\hline
exch2 & cd\_code & kpp & gfd & streamice & ptracers\\ 
\hline
gmredi & monitor & tapenade & exf & shelfice & thsice\\
\hline
cal & cost & ctrl & seaice & dic & grdchk\\
\hline
diagnostics & gchem & ggl90 & ecco & profiles & mnc\\
\hline
\end{tabular}}\\
\end{table}

The tangent linear and adjoint models are the linearized forward model and its transpose. The adjoint provides sensitivity information that may deviate from the response of the parent nonlinear forward model due to (i) the linearity assumption of the adjoint, (ii) inexactness, or (iii) incorrectness. Here we define \textit{inexactness} as deliberate omissions/approximations of problematic code structures in the nonlinear forward model to enable calculation of the adjoint. We distinguish this from adjoint \textit{incorrectness}, a truer error arising with flow reversal and detectable as a difference between tangent linear and adjoint gradients.

It is important to perform rigorous checks to ensure the adjoint information is meaningful \citep[e.g.,][Appendix A]{Loose2020}. Linearity checks consist of applying perturbations of the same (small) amplitude but opposite sign, $\pm \Delta u$, in the parent forward model. These yield approximately equal and opposite responses in an objective function, $\Delta^+_\text{fwd}\mathcal{J}\approx -\Delta^-_\text{fwd}\mathcal{J}$, if the dynamics are predominantly linear, which is more likely for short time periods and small perturbations. We here do not consider further details of numerical aspects of finite-differencing schemes. Exactness can be investigated (but not definitively identified in the presence of nonlinearity or incorrectness) by comparing these forward model responses to that given by convolving the applied anomalies with the adjoint-based sensitivities: 
\begin{equation}
\Delta^\pm_\text{adj}\mathcal{J}\,=\,\pm \Delta u\frac{\partial \mathcal{J}}{\partial u}\label{eqn:fd_grad}
\end{equation}
If the response is linear but the forward response and adjoint estimates diverge, inexactness is indicated \citep[see][for an example]{Loose2020}. 

Here we focus on testing the correctness of the adjoint via comparison with the gradient obtained via (1) finite differences (quantifying aggregated error from all 3 origins), and (2) the tangent linear model (quantifying error from inexactness and incorrectness). A gradient check package is available within the MITgcm to facilitate adjoint verification. The difference between gradients should ideally be $<1\%$ to pass the check.

\section{Results from Regression Tests}
\label{section:example}

\noindent
We present Tapenade-generated results for the following regression tests and discuss their validation:
\code{global_with_exf},
\code{global_ocean.cs32x15}, and 
\code{tutorial_global_oce_biogeo} (referred to as  \code{global_oce_biogeo}) from MITgcm tag \code{checkpoint68u}, available at \url{https://github.com/MITgcm/MITgcm/releases/tag/checkpoint68u}.

\subsection{Timing analysis between Tapenade and TAF}
\noindent
Table \ref{table:TAF_comparison} shows a timing analysis study where the relative efficiency of TAF- and Tapenade-generated code is compared for the \code{global_with_exf} (100 timesteps, 50 model days), \code{global_ocean.cs32x15} (360 timesteps, 360 model days), and \code{global_oce_biogeo} (60 model timesteps, 30 model days) verification experiments.

\begin{table}[H] 
\small
\caption{Timing comparison for Tapenade (TAP) against TAF. Time in seconds. The
runs are performed on Intel Xeon CPU E5-2695 v3 nodes (2.30 GHz clock rate, 35.84 MB L3 cache, 63.3 GB memory). Checkpointing intervals were optimized for the TAF runs to hold the maximum number of timesteps in memory; multi-level checkpointing was required only for \code{global_ocean.cs32x15}.
}
\label{table:TAF_comparison}
\centering{
\begin{tabular}{|c|c|c|c|c|}
\hline
Experiment & Forward & TAF & TAP & TAP/TAF\\
\hline
{\footnotesize global\_with\_exf} & 4.0& 15.8 & 35.4 & 2.2\\
\hline
{\footnotesize global\_ocean.cs32x15} & 30.7 & 173.4 & 1112.2 & 6.4\\
\hline
{\footnotesize global\_oce\_biogeo} & 26.4& 53.4 & 165.0 & 3.1\\
\hline
\end{tabular}}
\end{table}

The Tapenade-based adjoint simulation currently is about 2-7 times slower than the TAF-based adjoint simulation.
Since actual differentiated computations with either tool are similar, we believe this difference comes from the strategy for making available \textit{required} variables (e.g., elements of the model state needed to evaluate derivatives of nonlinear expressions or control flows).
Whereas TAF primarily uses a recompute-all approach from user-defined restart points, Tapenade primarily uses a store-all approach, combined with checkpointing. In the end this results in different storage/recomputation trade-offs. TAF achieves its performance through targeted insertion of directives for storing (as opposed to recomputing) required variables. As a result, MITgcm's dynamical core has on the order of 350 such STORE directives, with another $~$600 directives in the different model packages. This results in highly tuned, albeit labor-intensive adjoint code performance tuning.
In contrast, we applied no performance tuning for the Tapenade adjoint, whose default strategy is to apply checkpointing, and therefore repeated execution, at each and every procedure call. In our view, this accounts for most of the extra run time of the Tapenade AD code. We recently experimented with directives that instruct Tapenade to \textit{not} apply checkpointing at selected procedure calls and obtained speed-up factors ranging from 2 to 3 on MITgcm test cases~\citep{hascoët2024profilingcheckpointingschedulesadjoint}.
In light of this, the difference in performance seen here without any further efficiency tuning of the Tapenade adjoint is a promising result. It vastly reduces the need for expert knowledge to generate efficient code, and the code already performs well without user-inserted directives.


\subsection{Validation of AD gradients with finite difference gradients}
\noindent
We compared the gradients computed by Tapenade against finite differences provided by the \lstinline{gradient check} package and against existing output computed by TAF. 
For all the verification experiments mentioned in Table \ref{table:compatible_verification_exps}, Tapenade shows excellent agreement with the finite difference (FD) gradient calculated using the central finite difference scheme (see equation \ref{eqn:fd_grad}). Table \ref{table:fd_checks} shows an example of the gradient validation of both TAF and Tapenade against the FD gradient for the \code{global_with_exf} verification experiment at 5 different points.

\begin{table}[H] 
\caption{Validating TAF and Tapenade (TAP) against FD gradient for experiment \code{global_with_exf}. Similar results were found for other experiments and tutorials.}
\label{table:fd_checks}
\centering{
\begin{tabular}{|c|c|c|c|}
\hline
Coord. (i,j,k,bi,bj) & FD & TAF error & TAP error \\
\hline
(4,8,1,1,1) & -0.01995 & 2.610E-8 & 1.443E-8\\
\hline
(5,8,1,1,1) & -0.01962 & 9.242E-9 & -1.152E-8\\
\hline
(6,8,1,1,1) & -0.01966 & 1.435E-8 & 5.876E-8\\
\hline
(7,8,1,1,1) & -0.02165 & -1.670E-9 & -4.357E-9\\
\hline
(8,8,1,1,1) & -0.02090 & 3.241E-8 & -3.801E-9\\
\hline
\end{tabular}}\\
\end{table}

We treat accuracy as the relative error to the FD gradient, which we consider as the ground truth value.
There is no observable trend for accuracy between TAF and Tapenade. Sometimes, Tapenade is more accurate than TAF, and vice versa.

\subsection{Visualizing the adjoint}

\begin{figure}[t!]
        \centering
        \vspace*{-4ex}
        \includegraphics[width=\linewidth]{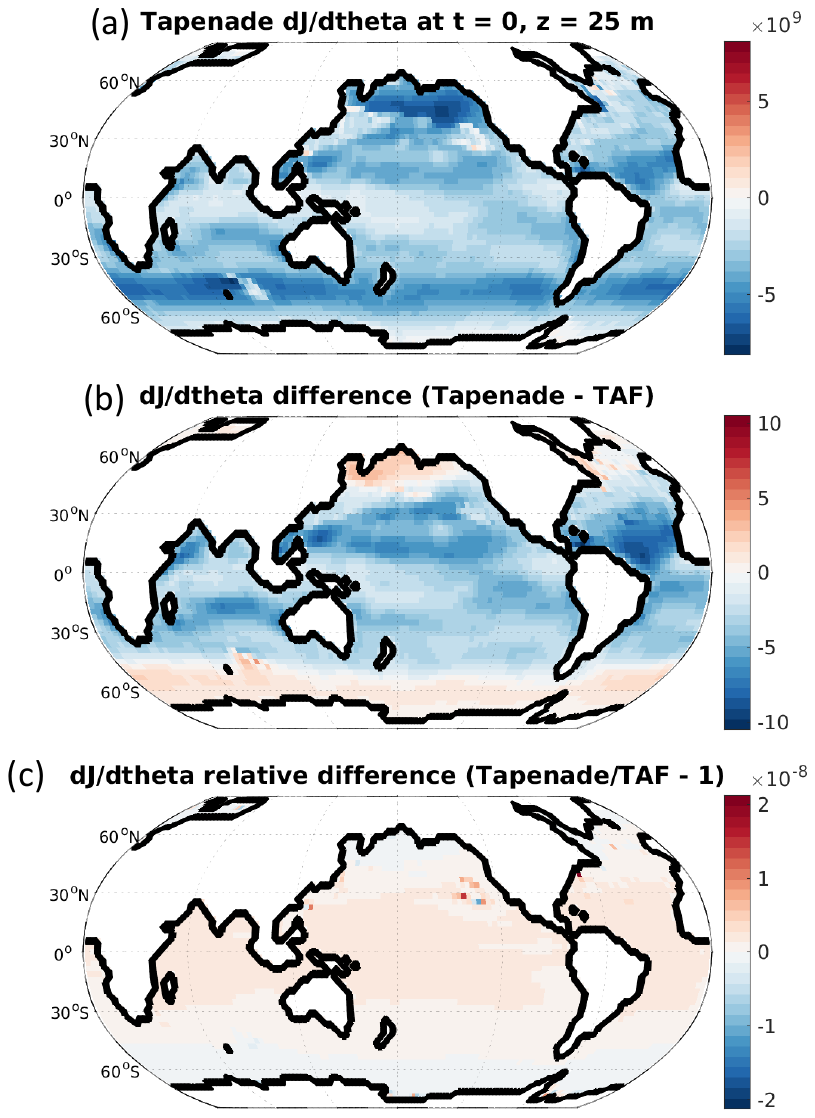}
        \vspace*{-2ex} 
        \caption{(a) Adjoint sensitivity (in mol/$^\circ$C) of the globally integrated air-sea flux of CO$_2$ on the final day of the integration to the initial SST in the experiment
        \code{tutorial_global_oce_biogeo}, computed using Tapenade-generated code. (b) The difference (in mol/$^\circ$C) between sensitivity maps generated with Tapenade- versus TAF-generated adjoint code. Note the difference in color scale between (a) and (b), indicating the strong quantitative consistency between sensitivities calculated with the two AD tools. Their pointwise differences are 0(10$^{8}$) times smaller than the sensitivity amplitude itself (panel c).}
        \label{fig:tut_biogeo}
        \end{figure}

\noindent
Having compared the performance and accuracy of the TAF- and Tapenade-based adjoint calculations, we now inspect the spatial structure and insights provided by the sensitivity distributions from the verification experiment \code{tutorial_global_oce_biogeo} (Fig. \ref{fig:tut_biogeo}). The dependent variable ($\mathcal{J}$) is the globally integrated air-sea flux of CO$_2$ on the final day of the integration. The sensitivity of $\mathcal{J}$ is plotted with respect to the initial sea surface temperature (SST) field (independent variable). Sensitivities derived using Tapenade (Fig. \ref{fig:tut_biogeo}a) and TAF show the same regional structure and strong quantitative consistency (Fig. \ref{fig:tut_biogeo}b \& c). Their pointwise differences are O(10$^{8}$) times smaller than the sensitivity amplitude. Widespread negative sensitivity indicates that an SST increase at almost any location leads to a decrease in globally integrated CO$_2$ uptake 1 month later. This is consistent with the established effect of warming on reducing CO$_2$ solubility in seawater. Strongest sensitivities are found between 40$^\circ$ and 60$^\circ$ latitude in both hemispheres, coincident with the location of strongest ocean CO$_2$ uptake in the modeled climatology and observations \citep{takahashi02}. These regions host both strong surface cooling of poleward-flowing waters and strong wind-driven nutrient upwelling and biological activity, facilitating CO$_2$ drawdown. The strong westerly winds in these latitude bands also invigorate air-sea gas exchange, further enhancing ocean uptake of CO$_2$ \citep{takahashi02}. 

In summary, the adjoint sensitivity distributions indicate that -- on the timescale considered -- surface cooling over regions primed for CO$_2$ uptake reinforces the ocean sink of CO$_2$. We anticipate that on longer timescales sensitivity distributions could reveal more complex circulation feedbacks \citep[e.g.,][]{deVries17} and this merits further investigation. For now, our example serves to demonstrate the rich and quantitatively consistent dynamical insights provided by the TAF- and Tapenade-generated adjoints.

\section{Conclusion \& Future Directions}\label{section:conclusion}

\noindent
The MITgcm, a widely used, state-of-the-art numerical model for simulating atmosphere and ocean circulation, has been enhanced with a new, open-source automatic-differentiation capability for generating efficient tangent-linear and adjoint code. This differentiable programming framework, \textit{MITgcm-AD v2}, is a powerful tool for data assimilation, sensitivity investigation, and dynamical attribution studies since it generates derivative code by employing source-to-source automatic/algorithmic differentiation with the open-source tool Tapenade.
The purpose of pushing for this open-source development was (i) to make ocean and climate model applications requiring comprehensive gradient information available to a much wider user community, (ii) to increase the community of developers contributing to the AD tool itself, and (iii) to enable comparison of adjoint simulations obtained from two completely different AD tools as another way of validating the gradients computed; ultimately, it is hoped that similar comparisons will become possible using the adjoint of different ocean models.
For the near future, we are prioritizing the following directions:
\begin{itemize}
    \item Integrate more completely with the in-built diagnostics package in the MITgcm to aid users in better managing their AD output, i.e., the (time-varying) adjoint fields (or dual state) produced by the adjoint model.
    \item Introduce writing and reading tape to files with Tapenade, just like TAF can do currently. This will allow Tapenade to handle larger simulations.
    \item Improve the performance of Tapenade-generated adjoint code. Current performance issues are expected to be related to  
    excessive systematic checkpointing at each subroutine call, leading to excessive recomputation.
    Looking at section 3.4.2 of \cite{Heimbach2005} helps identify possible performance bottlenecks due to contributions from computation, storage, and exchange. The computation of the adjoint model is more expensive than the computation of the forward model because it involves accumulation, variable resetting, and extra computation due to nonlinearity (product rule generating multiple terms). It also includes recomputation. Tapenade may not have as efficient recomputation algorithms as used by TAF. TAF achieves this through heavy use of STORE directives in the model. Such directives are not required by Tapenade.  
    Finally, TAF-generated adjoint uses the same communication architecture as MITgcm, so this should not contribute to the adjoint slowdown.
    \item Explore Tapenade enhancements to handle adjoint dump and restarts. This is currently enabled via divided adjoint (DIVA) within TAF and allows for long adjoint integrations to be performed despite typical wall time/memory limits on various HPC platforms.
    \item Get Tapenade to work more efficiently with some packages, for example, the specialized \code{streamice} package that uses the MITgcm dynamical core to perform ice sheet simulations \citep{Goldberg.2013}.
    \item Handle the out-of-bound issues observed with the \code{-devel} flag and NaNs observed with the \code{-ieee} flag.
    \item Make changes to polish the use of Tapenade through a docker, for MacOS users.
\end{itemize}
The long-term goal is to make open-source AD using Tapenade fully compatible with the ECCO State Estimation framework and thus make the adjoint-based data assimilation framework itself more accessible to a wide user base. Users can then choose their own model configuration, domain of interest, model-date misfit cost functions, set of control variables, period of integration, etc., for their science applications.
Open-source AD is also opening new avenues in the use of differentiable programming for Earth system models for existing Fortran-based models commonly used to date in the climate modeling community.
        
\section{Acknowledgements}
\noindent
SSG, HP, AN, and PH were supported in part via NSF \#1903596 and ECCO (NASA PO via subcontract to JPL/Caltech).
This material is based upon work supported in part by the Applied Mathematics activity within the U.S. Department of Energy, Office of Science, Advanced Scientific Computing Research Program, under contract number DE-AC02-06CH11357.

\appendix

\section{Reproducing Results}
\label{section:build_ad}

\noindent
This is a brief description of reproducing the results and plots in Section \ref{section:example}. All experiments require a similar procedure, so without loss of generality, we describe the steps for the verification experiment \code{tutorial_global_oce_biogeo} running on a Linux platform.

\subsection{Building the MITgcm}
\label{subsection:building_mitgcm}

    \noindent
    There are multiple ways to build the MITgcm, as described in section 3.2 of the MITgcm documentation (\url{https://mitgcm.readthedocs.io/en/latest/index.html}). We use \code{git} here to get the latest MITgcm tag, \code{checkpoint68u}.

\begin{lstlisting}[language = bash, rulecolor=\color{white}]
$ git clone https://github.com/MITgcm/MITgcm.git
\end{lstlisting}

    We then get the latest tagged release, \code{checkpoint68u} to ensure the reproducibility of results in the future.

\begin{lstlisting}[language = bash, rulecolor=\color{white}]
$ git checkout checkpoint68u
\end{lstlisting}

\subsection{Installing Tapenade}
\label{subsubsection:install_tapenade}

\noindent
    We detail below the instructions for Linux, but the latest instructions for many operating systems can always be found at: \url{https://tapenade.gitlabpages.inria.fr/tapenade/distrib/README.html}.
Before installing Tapenade, you must check that an up-to-date Java Runtime Environment is installed. Tapenade will not run with old Java Runtime Environments. Also, read the Tapenade license at \url{https://tapenade.gitlabpages.inria.fr/userdoc/build/html/LICENSE.html}.

\begin{enumerate}

\item Download the code from \url{https://tapenade.gitlabpages.inria.fr/tapenade/distrib/tapenade_3.16.tar} into your chosen installation directory \verb|install_dir|.

\item Go to your chosen installation directory \\\verb|install_dir|, and extract Tapenade from the tar file : 

\begin{lstlisting}[language = bash, rulecolor=\color{white}]
$ tar xvfz tapenade_3.16.tar
\end{lstlisting}

\item On Linux, depending on your distribution, Tapenade may require you to set the shell variable \\\verb|JAVA_HOME| to your Java installation directory. It is often \verb|JAVA_HOME=/usr/java/default|. You might also need to modify the \verb|PATH| by adding the bin directory from the Tapenade installation. Every time you wish to use the AD capability with Tapenade, you must re-source the environment. We recommend that this be done automatically in your bash profile upon login.

\end{enumerate}

You should now have a working copy of Tapenade. For more information on the \code{tapenade} command and its arguments, type :

\begin{lstlisting}[language = bash, rulecolor=\color{white}]
$ tapenade -?
\end{lstlisting}

\subsection{Running MITgcm-AD v2}
\label{subsection:running}

\noindent
Running an adjoint simulation with MITgcm using Tapenade is not very different from running any other typical MITgcm simulation and broadly involves the same steps.

\begin{itemize}

    \item Go to the build directory of the \code{tutorial_global_oce_biogeo} verification experiment. From the root of the MITgcm repository, this is the path:

\begin{lstlisting}[language = bash, rulecolor=\color{white}]
$ cd verification/tutorial_global_oce_biogeo/build
\end{lstlisting}

    \item Change the \code{nTimeSteps} parameter in \code{input_ad/data} file based on how long you want the simulation to be.

    \item Clean any previous remnant files in your build subdirectory from a previous simulation, if any.

\begin{lstlisting}[language = bash, rulecolor=\color{white}]
$ make CLEAN
\end{lstlisting}
    
    \item Generate a makefile using the \code{genmake2} script in the MITgcm. This command is run from the build subdirectory of your setup. The \code{-tap} option specifies that this is a Tapenade-based adjoint simulation. The \code{code_tap} directory refers to the Tapenade-equivalent of the \code{code_ad} directory used with TAF. Note that the optfile (option \code{-of}) can change based on the OS and hardware.

\begin{lstlisting}[language = bash]
$ ../../../tools/genmake2 -tap -of ../../../tools/build_options/linux_amd64_ifort -mods ../code_tap
\end{lstlisting}

    \item Run the make commands. \code{tap_adj} is the target name for the Tapenade adjoint mode and \code{tap_tlm} is the target name for the Tapenade tangent-linear mode. The results in Section \ref{section:example} are derived using the adjoint mode.

\begin{lstlisting}[language = bash]
$ make depend
$ make tap_adj
\end{lstlisting}

    These commands will generate the adjoint/TLM executable \code{mitgcmuv_tap_[adj,tlm]}.

    In the future, when Tapenade is integrated with MITgcm diagnostics, one will be able to simply rely on these diagnostics to get the relevant fields such as temperature output in either binary or NetCDF format. Currently, one must instead do the I/O manually as shown below.

    \item Open the Tapenade-generated file \code{the_main_loop_b.f} which is a differentiated and preprocessed version of the original MITgcm file \code{model/src/the_main_loop.F} and then add the following lines for I/O at the very end of the subroutine after all the \code{DO} loops but before the calls \verb|CALL POPREAL8ARRAY(theta, ...)| and \\\verb|CALL POPREAL8ARRAY(theta, ...)| since these are calls related to the Tapenade tape and will erase the values in these fields (remember, the adjoint code goes in the reverse direction so the last values of these fields will be zeros).

\begin{lstlisting}[language = fortran]
open(unit=500, file='hFacC_biogeo.data')
open(unit=501, file='theta_biogeo.data')
open(unit=502, file='salt_biogeo.data')
  DO ii1=1,nsy
  DO ii2=1,nsx
  DO ii3=1,nr
  DO ii4=1-oly,oly+sny
  DO ii5=1-olx,olx+snx
    write(500,*) hFacC(ii5,ii4,ii3,ii2,ii1)
    write(501,*) theta(ii5,ii4,ii3,ii2,ii1)
    write(502,*) salt(ii5,ii4,ii3,ii2,ii1)
  ENDDO
  ENDDO
  ENDDO
  ENDDO
  ENDDO
close(500)
close(501)
close(502)
\end{lstlisting}

    Then add the following lines at the very end of the subroutine to get the adjoint values or the gradient with respect to the initial values of \code{salt, theta}.

\begin{lstlisting}[language = fortran]
open(unit=503, file='thetab_biogeo.data')
open(unit=504, file='saltb_biogeo.data')
  DO ii1=1,nsy
  DO ii2=1,nsx
  DO ii3=1,nr
  DO ii4=1-oly,oly+sny
  DO ii5=1-olx,olx+snx
    write(503,*) thetab(ii5,ii4,ii3,ii2,ii1)
    write(504,*) saltb(ii5,ii4,ii3,ii2,ii1)
  ENDDO
  ENDDO
  ENDDO
  ENDDO
  ENDDO
close(503)
close(504)
\end{lstlisting}

    \item Copy this manually edited \code{the_main_loop_b.f} file for safekeeping and reuse since re-running the Tapenade command will generate a fresh file without the manual changes.

\begin{lstlisting}[language = bash, rulecolor=\color{white}]
$ cp the_main_loop_b.f ..
\end{lstlisting}

    \item In \code{genmake2}, add the following line to the steps for target \code{TAP_ADJ_FILES}. This will copy the safely kept, manually edited file \code{the_main_loop_b.f} back to the build directory \emph{after} the Tapenade command is run and before the final compilation so the manual changes have the desired effect.

\begin{lstlisting}[language = bash, rulecolor=\color{white}]
$ cp ../the_main_loop_b.f .
\end{lstlisting}

    \item Linking the input files and the executable in the run directory. This is a standard step in the MITgcm. It is performed from the run directory.

\begin{lstlisting}[language = bash]
$ cd ../run
$ rm -r *
$ ln -s ../input_tap/* .
$ ../input_tap/prepare_run
$ ln -s ../build/mitgcmuv_tap_adj .
\end{lstlisting}

    \item Running the executable and keeping logs.

\begin{lstlisting}[language = bash, rulecolor=\color{white}]
$ ./mitgcmuv_tap_adj > output_tap_adj.txt 2>&1
\end{lstlisting}

    \item One can analyze and plot the 2D fields after reading the output files, for example, \code{thetab_biogeo.data} and running it through the following Python commands:

\begin{lstlisting}[language = python]
thetab = np.loadtxt(thetab_biogeo.data)
# New shape from code_tap/SIZE.h
thetab = np.reshape(thetab, (2,2,15,40,72))
# Get rid of the halo points
thetab = thetab[:, :, z_level, 4:-4, 4:-4]
thetab_globe = np.zeros((64,128), dtype = float)
thetab_globe[:32,:64] = thetab[0,0,:,:]
thetab_globe[32:,:64] = thetab[1,0,:,:]
thetab_globe[:32,64:] = thetab[0,1,:,:]
thetab_globe[32:,64:] = thetab[1,1,:,:]
\end{lstlisting}

    \code{thetab_globe} has the shape \code{(64,128)}.

    \item Get the water-land mask for the surface in a similar fashion by reading in \code{hFacC_biogeo.data} and running it through the same commands as above.

\begin{lstlisting}[language = python]
hfacc = np.loadtxt(hFacC_biogeo.data)
# New shape from code_tap/SIZE.h
hfacc = np.reshape(hfacc, (2,2,15,40,72))
# Get rid of the halo points
hfacc = hfacc[:, :, z_level, 4:-4, 4:-4]
hfacc_globe = np.zeros((64,128), dtype = float)
hfacc_globe[:32,:64] = hfacc[0,0,:,:]
hfacc_globe[32:,:64] = hfacc[1,0,:,:]
hfacc_globe[:32,64:] = hfacc[0,1,:,:]
hfacc_globe[32:,64:] = hfacc[1,1,:,:]
\end{lstlisting}

    \code{hfacc_globe} also has the shape \code{(64,128)}.
    \item The filled contour plots in Section \ref{section:example} are then created by plotting \code{hfacc_globe*thetab_globe} to account for the land mask while plotting the adjoint of the temperature field.
    
\end{itemize}

 \bibliographystyle{elsarticle-num} 
 \bibliography{jlesc,refs}

\pagebreak
\onecolumn
\framebox{\parbox{\columnwidth}{The submitted manuscript has been created by UChicago Argonne, LLC, Operator of Argonne National Laboratory (`Argonne'). Argonne, a U.S. Department of Energy Office of Science laboratory, is operated under Contract No. DE-AC02-06CH11357. The U.S. Government retains for itself, and others acting on its behalf, a paid-up nonexclusive, irrevocable worldwide license in said article to reproduce, prepare derivative works, distribute copies to the public, and perform publicly and display publicly, by or on behalf of the Government.  The Department of Energy will provide public access to these results of federally sponsored research in accordance with the DOE Public Access Plan. \url{http://energy.gov/downloads/doe-public-access-plan}.}}

\end{document}